\documentclass[letterpaper,showpacs,pre,superscriptaddress,twocolumn,floatfix]{revtex4}
\usepackage{graphicx}
\usepackage{psfrag}
\usepackage{color}
\usepackage{amsbsy}
\usepackage{amssymb}

\def\e{\varepsilon}

\begin{document}

\DeclareGraphicsExtensions{.pdf,.jpg,.tif} 
     

\title{First passage times of pulling-assisted 
DNA unzipping}

\author{G. Lakatos}
\affiliation{Departments of Physics and Chemistry, The University of British Columbia, Vancouver, BC, Canada, V6T-1Z1}
\author{T. Chou}
\affiliation{Department of Biomathematics, UCLA, Los Angeles, CA 90095-1766}
\author{B. Bergersen}
\affiliation{Department of Physics, The University of British Columbia}
\author{G. N. Patey}
\affiliation{Department of Chemistry, The University of British Columbia}

\begin{abstract}

We investigate the voltage-driven transport of hybridized DNA through
membrane channels.  As membrane channels are typically too narrow to
accommodate hybridized DNA, the dehybridization of the DNA is the critical
rate limiting step in the transport process.  Using a two-dimensional stochastic model,
we show that the dehybridization process proceeds by two 
distinct mechanisms; thermal denaturation in the limit
of low driving voltage, and direct stripping in the high to moderate voltage
regime.  Additionally, we investigate the effects of introducing non-homologous defects
into the DNA strand.


\end{abstract}



\maketitle

\section{Introduction}


Over the last two decades, the rapid development of single molecule
manipulation techniques has produced dramatic advances in chemistry and molecular 
physics.  The fields of DNA biotechnology and nanofabrication in particular, 
have benefited from these advances.  In turn, successes in the field of DNA biotechnology have
further motivated the study of DNA manipulation techniques, with a strong
focus on studying mechanisms of inducing DNA dehybridization (\cite{LUBENSKY1}, \cite{COCCO1}).

In this study, we consider dehybridizing DNA 
by pulling DNA molecules through a 
transmembrane channel.  This work is motivated in part by the 
experimental results of Nakane {\it et al.} \cite{NAKANE},
and Bates {\it el al.} \cite{BATES}. In the experiments of 
Nakane, a probe DNA that extends 14 base pairs beyond 
the opening of a trans-membrane channel is
used to capture a 14 base pair target strand (see Fig \ref{FIG1}).  In order
to prevent complete passage of the probe DNA through the membrane channel,
the probe was attached to an Avidin anchor protein by a 50-base poly-A tail.
The target strands used in the experiments were either completely homologous to 
the probe strand, or had one 
nonhomologous ``defect" base.  After the probe strand had captured
a target strand, a transmembrane voltage was applied to 
pull the probe strand through an $\alpha$-hemolysin channel embedded in
a lipid membrane.  As the $\alpha$-hemolysin channel could only accommodate
{\it single-stranded} DNA molecules, any target strand hybridized with the
probe DNA would need to detach completely before the probe could be fully drawn
into the channel.  As the transmembrane voltage was applied, the ionic current
through the $\alpha$-hemolysin channel was simultaneously 
monitored. When any part of the probe strand was inside the channel, the channel was blocked,
and the ionic current fell below the measurement threshold.  When the probe strand was 
completely pulled through the channel (after the target strand detached), 
ionic currents resumed and were measured. Thus, in the experiment of Nakane, 
the distribution of first passage times, $\mathrm{   }\tau$, for the
escape of the probe DNA from the $\alpha$-hemolysin channel was measured.  

In order to understand the first passage times observed
in these experiments, we have produced estimates for the thermally averaged 
mean first passage time $(\tau_\mathrm{{eq}})$ using simple
two-dimensional and one-dimensional stochastic models.  
We model the DNA hybridization energetics using
both a simple free energy model where each hybridized base contributes an equal amount
to the total free energy, and using free energies produced by the MFOLD 2-state
hybridization server \cite{MFOLD}.  Our predicted mean first passage times
show the same qualitative features observed by Nakane {\it et al.}  Under the influence
of small transmembrane voltages, the first passage times are found to depend strongly on
the presence and energy of defects in the target DNA strand.  
As the transmembrane voltage is increased, a distinct roll-over in the predicted
first passage times is observed, and the first passage times become relatively insensitive to
the presence of defects on the target DNA strand.  We propose that this roll-over
in $\tau_\mathrm{{eq}}$ represents a transition between a thermally-dominated dehybridization
mechanism in the low voltage regime, and
a driven stripping mechanism in the high voltage regime.  


\section{MODEL}

\begin{figure}
  \includegraphics[width=3.3in]{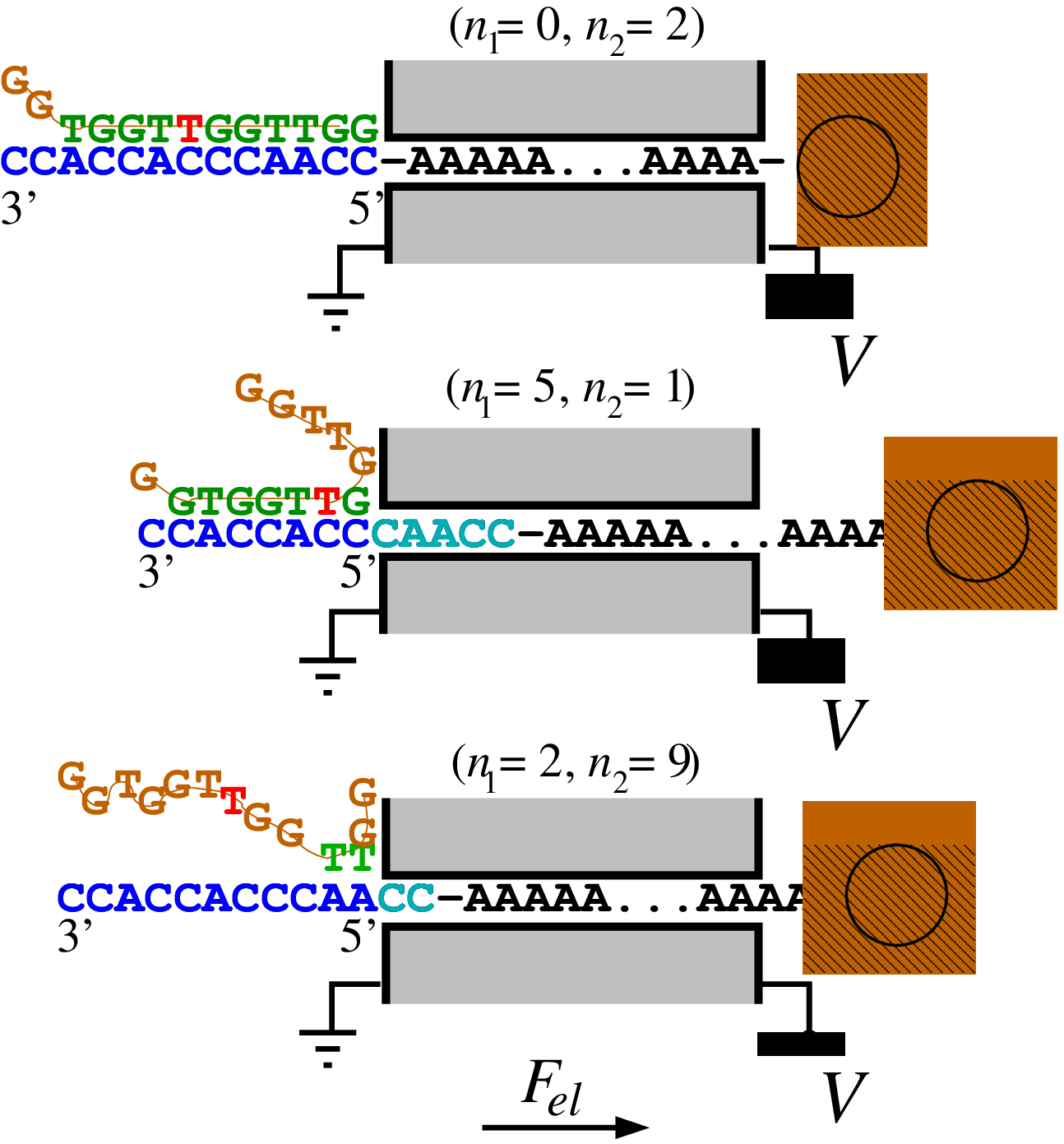}
  \caption{An example of the DNA detachment experiment. 
The detachment of the length $N$ target DNA is assumed 
to occur via two coordinates $n_{1}$ and $n_{2}$, 
representing unzipping from the two ends of the probe DNA. Here, an $N=14$ 
target strand is shown with a single defect.  The
probe DNA is shown in blue, hybridized target DNA in green,
the defect base in red, dehybridized bases in brown, and
extracted bases in light blue.  The probability of bubbles forming is ignored. }
  \label{FIG1}
\end{figure}

Motivated by the double-stranded DNA unzipping models of Cocco {\it et. al.} \cite{COCCO1} and
Poland and Scheraga \cite{POLAND}, we model the process of DNA probe 
extraction using a two-sided zipper model described in Fig. \ref{FIG1}.  
This model generates $\frac{(N+1)(N+2)}{2}$ 
distinct partial dehybridization states labeled $(n_1,n_2)$, 
were $n_1$ is the number of bases at the $5'$
end of the probe that are dehybridized {\it and} have been 
pulled into the transmembrane region,
and $n_2$ is the number of dehybridized bases 
at the $3'$ end of the DNA probe. Using this nomenclature, the
DNA probe is fully dehybridized when $n_1 + n_2 = N$, 
and fully {\it extracted} only when $n_1 = N$ 
and $n_2 = 0$.  Note that his model neglects the effects of
internal dehybridization bubbles, which will be discussed later
in this section.  Defining $P_{(n_1,n_2)}(t)$ 
to be the probability of being in state $(n_1,n_2)$ 
at time $t$, given that the system was 
initially in state $(n'_1,n'_2)$,
we find the system evolves according to a master equation

\begin{equation}
\displaystyle \frac{d\vec{P}(t)}{dt} = {\bf M}\vec{P}(t).
\label{EQN1}
\end{equation}

\noindent In order to produce the elements $M_{i,j}$ of the 
transition matrix, we define the 
following five transition mechanisms between states $(n_1,n_2)$:

\begin{description}

\item[$\bullet \mathrm{ } 5'$ Bond Breaking] \label{RATE3}  The hydrogen 
bonds between bases at the $5'$ end of 
the probe DNA break, the probe DNA base moves into 
the transmembrane region, and the state 
shifts from $(n_1,n_2)$ to $(n_1 + 1, n_2)$.  
This transition path includes
a bond breaking event, immediately followed by translation
of the probe DNA, and occurs at rate

\begin{equation}
k_1 = \mu_1 e^{- \beta (\Delta G + \Delta V)}
\label{BONDBREAKRATE1}
\end{equation}

Where $\Delta G$ is the change in free energy between the 
current state $(n_1, n_2)$ and the 
target state $(n_1+1, n_2)$, and 
$\mu_1$ is a phenomenological attempt frequency.  We note
that we could express $\mu_1$ as the product of an attempt
frequency and a free energy barrier $\mu_1 \rightarrow \mu e^{- \Delta G^{*}/kT}$.
However for the $5'$ bond-breaking transition, as well as all other 
transitions in our model, we pull any dependence on the barrier energy
into the effective attempt frequency.
As this state transition involves a net translation of 
a partially charged nucleotide across the membrane, a drop in 
the interaction energy between the applied electric field and the 
probe DNA occurs: 

\begin{equation}
\label{RATEEQ2}
\Delta V = V(n_1 + 1,n_2) - V(n_1,n_2) = q V,
\end{equation}


where $q$ is the mean partial charge of a DNA base.


\item[$\bullet \mathrm{ } 5'$ Bond Formation] \label{RATE4}  In the reverse of the 
$5'$ bond breaking transition, the probe DNA
translates one base in the $3'$ direction, and corresponding 
bases on the probe and the target DNA
strands hybridize.  This shifts the system from the 
$(n_1,n_2)$ state to the  $(n_1 - 1, n_2)$ state, and occurs at rate

\begin{equation}
r_1 = \nu_1 e^{-\beta (\Delta V)}
\end{equation}

\item[$\bullet \mathrm{ } 3'$ Bond Breaking] \label{RATE1}
The hydrogen bonds between two hybridized bases located 
at the $3'$ end of the probe break, thus increasing $n_2$ by one.  
This transition is assumed to occur with rate

\begin{equation}
k_2 = \mu_2 e^{-\beta (\Delta G)}
\label{BONDBREAKRATE2}
\end{equation}

\item[$\bullet 3'$ Bond Formation] \label{RATE2} Corresponding 
dehybridized bases at the $3'$ end of the 
probe strand and the $5'$ end of the target DNA strand form a bond, 
decreasing $n_2$ by one.  This transition is
assumed to occur at rate

\begin{equation}
r_2 = \nu_2
\end{equation}

\item[$\bullet \mathrm{ }$ Free Translation] \label{RATE5} In the case where the 
DNA probe is completely dehybridized
$(n_1 + n_2 = N)$, but not fully extracted $(n_1 \neq N)$, 
the DNA probe will translate under the influence of the applied potential.  
The translation changes the state from $(n_1,n_2)$ 
to $(n_1 + 1, n_2 - 1)$ or $(n_1 - 1,n_2 + 1)$ and occurs with rate

\begin{equation}
\begin{array}{ccc}
k_{3} &=& \mu_3 e^{-\beta (\Delta V)}\\
k_{4} &=& \mu_3 e^{-\beta (\Delta V)}\\
\end{array}
\end{equation}

The two rates $k_3$ and $k_4$ describe translation in the $5'$ and
$3'$ directions respectively.  The only significant difference
between $k_3$ and $k_4$ is the sign of the change in the interaction
energy between the probe DNA and the applied electric field.
In the case of $5'$ translation $\Delta V < 0$, while
in the case of $3'$ translation $\Delta V > 0$.
\end{description}

Having defined the possible transitions in the model, we are left with
five unknown attempt frequencies $(\mu_1,\nu_1,\mu_2,\nu_2,\mu_3)$, and 
an unknown value for the mean partial charge per DNA base $q$.
Based on the results of Nakane, we set $q = 0.4 e$ \cite{NAKANE},
were $e$ is the electron charge.  Since free translation and $5'$ bond breaking 
both involve sliding of the
probe DNA strand, the time scale for these transitions is set approximately by the
time scale for translation of the single-stranded probe DNA.
To find this time scale, we built a simplified model of 
single-stranded DNA translation consisting of only
the free translation moves described above.  Using this model, we 
simulated the complete passage of a 60-base single-stranded DNA segment through a channel
30 DNA bases in length.
We then compared the mean first passage time for complete transit of single-stranded DNA computed
from this model, with the experimental results of Bates {\it et al} \cite{BATES} for DNA
transport through $\alpha$-hemolysin channels at low voltages ($1010 \mu \mathrm{s}$ at 20mV,
$530 \mu \mathrm{s}$ at 40mV).  The best match between the computed and experimental results
occurred when the attempt frequency was approximately $\mu_1 = 9 \times 10^5\mathrm{s}^{-1}$.
To determine a value for $\nu_2$, we followed Cocco {\it et al.} \cite{COCCO1} and set
$\nu_2$ to equal the inverse self diffusion time for a single nucleotide 
($\approx 5\times10^6 \mathrm{s}^{-1}$).  With rough order of magnitude estimates
for $\mu_1$ and $\nu_2$ we then tuned the parameters of our model to give a reasonable
match to the experimental data of Nakane \cite{NAKANE} for the variation of escape time
with voltage.  After hand tuning, the parameters were $\mu_1 = 9\times10^5 \mathrm{s}^{-1}$,
$\mu_2 = 1.4\times10^7 \mathrm{s}^{-1}$, $\nu_1 = 3.5\times10^5 \mathrm{s}^{-1}$, 
$\nu_2 = 6\times10^6 \mathrm{s}^{-1}$, and $\mu_3 = 9\times10^5 \mathrm{s}^{-1}$.

While easy to analyze, our model fails
to capture a number of aspects of the denaturation and extraction
process.  For example, by assuming the short probe-target DNA
complex denatures only from the ends, we have ignored the
effects of bubble formation.  By bubbles 
we refer to dehybridized regions of DNA that are bounded by at least
one hybridized base on both sides.  While clearly an approximation,
we do not believe that bubble formation will qualitatively
affect the process of driven DNA hybridization described here.
Under physiologic
conditions, structural fluctuations in double stranded DNA lead to the formation of bubbles
that are typically on the order of several tens of base pairs
\cite{HANKE}.  In this paper we consider the dehybridization and
extraction of DNA strands only 10-20 bases in length, and thus
considerably smaller than the typical bubble size found in
DNA at physiologic conditions.  Additionally, recent experiments
\cite{ALTAN} have studied the formation of bubbles in small, 
atypically bubble-prone, poly-AT DNA chains approximately $18$ 
bases in length.  Here, the
formation of DNA bubbles approximately $2-10$ bases in length
were observed in the DNA strand.  However, through fluorescence-correlation
spectroscopy Altan-Bonnet \cite{ALTAN} were able to measure the
typical lifetime of these small bubbles to
be approximately 50$\mu s$.  This lifetime
is typically much smaller than the mean time for the dehybridization
and extraction of small DNA strands through $\alpha$-hemolysin at
moderate driving voltages.

Additionally we do not model partial-registry binding,
where the target DNA binds to the probe DNA with an
offset so that the target DNA ``over-hangs'' the probe.
If we assume that the target-probe DNA complex is
well equilibrated prior to the start of the extraction
processes, then such out of registry binding is unlikely
due to its high energetic cost.  Moreover, should
such binding occur, its effects on the DNA dehybridization
process would not be qualitatively different from the effects
of starting in a $(n_1,n_2)$ state with a similar number
of unhybridized bases.  

Finally, we do not model the
full extraction of the probe DNA ({\it i.e.,} full removal from the
$\alpha$-hemolysin channel).  As re-binding of the target DNA
is unlikely once the probe is fully drawn into the channel
(typical target concentrations are 10$\mu$M \cite{NAKANE}),
complete removal of the probe DNA depends primarily on the
details of the DNA-channel interactions.  Previous studies
\cite{BATES}, have shown that translation of single-stranded
DNA 60 bases in length under moderate driving (0.04V)
in an $\alpha$-hemolysin channel occurs on a time scale of
500$\mu$s.  Rather than model the interactions between the
probe DNA and channel, we can simply add $500 \mu \mathrm{s}$
to the first passage times for probe escape we compute
using our model.  As we shall see, at low to moderate
membrane voltages this additional time is small
compared to the time required to denature the probe
DNA.


\section{RESULTS AND DISCUSSION}

We use Eqn \ref{EQN1} to compute the thermally 
averaged time for the DNA probe to first reach 
the fully extracted state 
$(n_1 = N, n_2 = 0)$.  To do this, we make the $(N,0)$ 
state completely absorbing, and slightly modify Eqn \ref{EQN1} 
by setting $P_{(N,0)}(t) = 0$ for all $t$:

\begin{equation}
\begin{array}{ccccc}
\frac{\displaystyle d P_{i}}{\displaystyle dt} = \displaystyle \sum_{j \neq (N+1)} 
P_j(t) M_{i,j} + P_i(t) M_{i,i} & , & i \neq (N+1).\\
\end{array}
\label{FPT0}
\end{equation}

\noindent Note that the first term in Eqn \ref{FPT0} does not 
include contributions from transitions out
of state $(N,0)$ while the second term does include transitions
into state $(N,0)$.
Upon defining $S(t \vert (n_1',n_2'))$ 
to be the probability that the
system reaches the absorbing state between 
time $t$ and time $t + dt$, given that
it started in state $(n_1',n_2')$ we find

\begin{equation}
\begin{array}{ccc}
\displaystyle S(t \vert (n_1',n_2')) & = & 
\displaystyle -\frac{d}{dt} \sum_{j \neq (N+1)} P_{j}(t)\\
     & = & -\sum_{j \neq (N+1)} P_{j} M_{N+1,j}.\\
\end{array}
\label{FPT1}
\end{equation}

\noindent With $S(t \vert (n_1',n_2'))$ computed from the 
solution to Eqn \ref{FPT0}, we can determine the 
mean first passage time to the extracted state

\begin{equation}
\begin{array}{ccl}
\langle \tau(n_1',n_2')\rangle  & = & 
\displaystyle \int_{0}^{\infty} t S(t \vert (n_1',n_2')) dt \\
                & = & 
\displaystyle \int_{0}^{\infty}  \sum_{j \neq (N+1)} P_{j}(t) dt.\\
\end{array}
\label{FPT2}
\end{equation}

\noindent Finally, to find the thermally averaged mean first 
passage time, $\tau_\mathrm{{eq}}$, we sum the results of
Eqn \ref{FPT2} over all possible initial states of the probe DNA,

\begin{equation}
\tau_\mathrm{{eq}} = \frac{\sum_{n_1,n_2} \langle \tau(n_1,n_2)\rangle  
e^{-\beta G(n_1,n_2)}}{\sum_{n_1,n_2} e^{-\beta G(n_1,n_2)}}.
\label{FPT3}
\end{equation}

\noindent where $G(n_1,n_2)$ is the free energy of state $(n_1,n_2)$ 
relative to the free energy of the 
completely dehybridized system.  The validity of
Eqn \ref{FPT3} is contingent on the assumption of 
local thermal equilibrium between the probe and
the fluid bath.  However, as the probe is typically maintained 
in the inserted state for several tenths of
a second prior to removal \cite{NAKANE}, we expect this 
assumption to hold to reasonable accuracy.

\begin{figure}
  \includegraphics[width=3.3in]{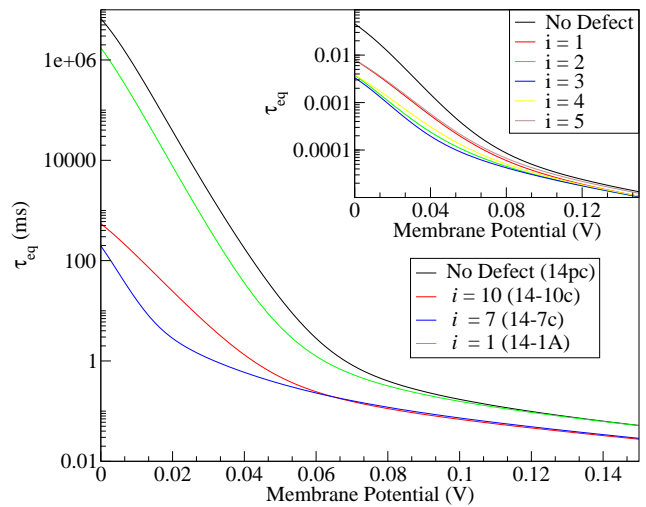}
  \caption{ Mean first passage times generated using free energies provided by the MFOLD program (see
Table \ref{SEQUENCES}).  At low voltages, energetic effects account for the majority of the 
difference in escape times between the 
various defect-carrying strands.  
INSET: Thermally averaged mean escape times of a
DNA segment 5 bases long. Each hybridized base lowers 
the free energy by $2 kT$, except for a single defect site 
(located at position $i$) which lowers the free energy by $kT/10$ when 
hybridized. Placing the defect in the center 
of the segment lowers the mean escape time significantly more than 
placing a defect at either end of the segment.}
  \label{LIFETIME1}
\end{figure}

To investigate how nonhomologous defects influence the 
mean first passage time, we first analyze the 
process of DNA extraction using a simple model where
each non-defective hybridized base pair lowered the free energy of the
probe-target complex by a constant amount.  A single defective
base pair with a binding free-energy lower than all other
bases was then introduced into the target strand.
As an example, we considered a short 
strand of eleven bases, and used the double zipper model to generate the
mean escape time curves displayed in the inset of Fig \ref{LIFETIME1}.  
There are two features worth noting; the mean escape 
time clearly depends on the location of a 
defective base as well as the energy of the defect, 
and the position and energy of the defect influence the escape time 
less as the applied voltage is increased. 
At first glance the dependence of the escape time on defect location 
appears counter-intuitive given that the defect energy 
did not vary with the location of the defect along the
target strand.  Nonetheless, the behavior 
displayed in Fig. \ref{LIFETIME1} where a defect placed in 
the center of the DNA strand produces 
maximal reduction in mean escape time, is extremely general and 
rather insensitive to changes in the 
parameter values of the model.

To understand the source of this effect, 
consider a DNA chain with a defect at position $i$,
when the DNA chain is in the $(i - 1,j)$ 
or $(j,i - 1)$ state with $j < (N - i + 1)$. 
If the defect is at site $i$,
the probability of the adjacent DNA base 
rehybridizing goes as $r/(r + k)$. Here $k$ is the
rate at which the defect base dehybridizes, 
and $r$ is the rate of rehybridization of the base
adjacent to the defect.  The effect of 
introducing a defect is to increase
$k$ while keeping $r$ constant, reducing the 
likelihood of hybridization of the adjacent
base.  In the case where the defect results in 
particularly weak binding, the increase in
$k$ can be quite large, making it highly unlikely 
that DNA will rehybridize beyond the
defect base $i$ once the defect itself has dehybridized.  
This shortens the effective length of the DNA by $\mathrm{min}(i-1,N -i + 1)$ sites.  

We note however that introducing bubbles into
the model would minimize the variation in escape time with defect position.
In the presence of bubbles, bases well separated from the defect could
rehybridize without the defect base itself rehybridizing.  This would dramatically 
reduce the variation in the mean escape time associated with the defect position.
Nonetheless, the mean escape time of GC-rich strands which are relatively
resistant to bubble formation may still be sensitive to the position of defects in the target strand.


\begin{table}
\begin{tabular}{|c|c|c|}
\hline
Name&Sequence&Energy\\
\hline \hline
14pc&$5'$-GGTGGTTGGTTTGGTT-$3'$&$-37.7kT$\\
1A&$5'$-GGTGGTTGGTTTGGT\underline{\textbf{A}}-$3'$&$-36.9kT$\\
10C&$5'$-GGTG\underline{\textbf{C}}TTGGTTTGGTT-$3'$&$-27.4kT$\\
7C&$5'$-GGTGGTT\underline{\textbf{C}}GTTTGGTT-$3'$&$-27.1kT$\\
\hline
\end{tabular}
\caption{DNA target sequences \cite{NAKANE} along with the hybridization energies predicted by
MFOLD ($T = 293$).  The probe DNA strand is perfectly complementary to the 14pc sequence.
The difference in the free energies between the various defect strands is due to the
presence of different defect nucleotides (\textbf{A} vs \textbf{C}), and differences
in the bases neighboring the defects.  For more detail see \cite{MFOLD}.}
\label{SEQUENCES}
\end{table}

Turning to a more realistic free energy model, we used the 
the MFOLD 2-state hybridization server \cite{MFOLD} to generate free energy 
surfaces for four distinct probe sequences, each 14 bases in length
\cite{NAKANE} (see Table \ref{SEQUENCES}).  The form of the lifetime verses voltage 
behavior produced by Eqn \ref{FPT3} for these
sequences is displayed in Fig. \ref{LIFETIME1}.  
The four voltage response curves clearly
separate into two groups; the 14pc sequence 
(perfectly complementary to the DNA probe) and
the 1A sequence (with a single non-complementary 
base at the $5'$ end of the sequence) 
belong to a long lived set, while the 7C and 10C 
sequences (both with a single non-complimentary
base in the center of the sequence) form a group with 
a dramatically shorter lifetime. We note that the 
approximately four orders of magnitude difference 
between the defect-free and 1A escape times and the 
7C and 10C escape times at zero voltage, 
is consistent with the free energy
differences between the various probes. 
From the standpoint of potential applications to DNA sequencing, 
the reduction in free energy produced by
a defective base is clearly the dominant influence 
on the mean escape time. 

Returning to the DNA sequences used to produce 
Fig \ref{LIFETIME1}, we investigated the
consequences on the mean probe lifetime of varying 
the $3'$ detachment rate, and found
that this rate has a significantly 
greater influence on the mean escape
time in the low voltage regime than in the high voltage regime. 
This leads us to propose that the field-induced dehybridization
can occur through one of two mechanisms.  In the low voltage regime, thermal 
denaturation of the target DNA from the probe
is the dominant mode of separation.  In this region 
of the lifetime-voltage curve, thermally-induced 
dehybridization from the $3'$ end is at least as 
significant as voltage driven-bond
stripping at the $5'$ end.  As the applied voltage 
is increased, the system enters a transition
region where voltage-driven bond stripping becomes 
the statistically favored mechanism of
detachment.  Finally, in the high voltage region 
stripping becomes the primary means of separation.

To confirm this hypothesis, we find 
the probability that base $i$ (measured from
the $5'$ end of the probe DNA), is the last base 
to dehybridize prior to extraction of the DNA
probe.  We first define $g_{(i,(N-1) - i)}(t) = g_{i}(t)$
to be the probability density that the last
dehybridization event occurs at base $i$ between time $t$ and
$t + dt$.  Making use of the solution to Eqn \ref{EQN1}, $g_{i}(t)$
can be expressed as

\begin{equation}
\label{PATHPROB0}
g_{i}(t) = P_{i}(t) k_{2}^{(i)} Q_{i}(t,t^{*}) + P_{i}(t) k_{1}^{(i)} Q_{i +1}(t,t^{*}).
\end{equation}

\noindent Here $Q_{i}(t,t^{*})$ is the probability the probe DNA is fully drawn into
the channel by time $t^{*}$ given that it was fully dehybridized ({\it i.e.,} in state $(i,N - i)$)
at time $t$, assuming that no rehybridization events occur between $t$ and $t^{*}$.  Rates
$k_{1}^{(i)}$ and $k_{2}^{(i)}$ are defined by Eqns \ref{BONDBREAKRATE1} and \ref{BONDBREAKRATE2} respectively.

Since the probabilities produced by Eqn \ref{EQN1} contain contributions
from paths through the $(n_1,n_2)$ state space that 
include rehybridization events, we cannot
use the solution to Eqn \ref{EQN1} to compute $Q_{i}$.  
However we can compute $Q_i$ by considering the
master equation governing the transitions between
fully dehybridized states $(n_1,N-n_1)$.  Defining 
$W_{(i,N-i)}(t;(j,N-j))$ to be the probability of
being in the $(i,N-i)$ state at time $t$, given that the system was in
state $(j,N-j)$ at time $t = 0$, we have

\begin{equation}
  \begin{array}{ccll}
     \displaystyle \frac{dW_{(0,N)}}{dt} & = & -(r_1^{(0)} + k_3^{(0)})W_{(0,N)} + k_4^{(0)} W_{(1,N-1)} & \\[13pt]
     \displaystyle \frac{dW_{(i,N-i)}}{dt} & = & -(r_2^{(i)} + r_1^{(i)} + k_3^{(i)} + k_4^{(i)})W_{(i,N-i)} &\\[13pt]
     & & + k_3^{(i-1)}W_{(i-1,N-(i-1))} + k_4^{(i+1)}W_{(i+1,N-(i+1))} &\\[13pt]
     & & 0 < i < (N-1)  & \\[13pt]
     \displaystyle \frac{dW_{(N-1,1)}}{dt} & = &-(r_2^{(N-1)} + r_1^{(N-1)} + k_3^{(N-1)} + k_4^{(N-1)})W_{(N-1,1)}\\[13pt]
     \displaystyle & & + k_3^{(N-2)}W_{(N-2,2)}. &\\[13pt]
  \end{array}
  \label{AUXMASTER1}
\end{equation}

\begin{figure}
  \includegraphics[width=3.3in]{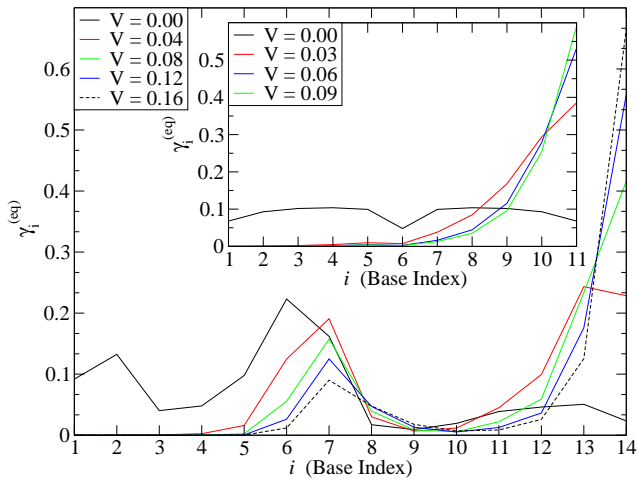}
  \caption{The last bond probabilities for the 
10c probe at various voltages.  At low
  voltages the distribution is determined primarily
by the free energy surface 
  indicating that thermally-driven dehybridization is 
the dominant dehybridization mechanism.  With increasing 
voltage, the effects of the free energy surface
  are rapidly obscured and stripping is the dominant 
denaturation mechanism.  INSET: The last bond probabilities 
for an 11 base DNA with a defect
  at position $i = 5$.  The configurational 
free energies were produced assuming
  a simple free energy model with a uniform free energy per bond,
  and uniform attempt frequencies of $10^6 \mathrm{s}^{-1}$.}
  \label{BONDBREAK1}
\end{figure}

In Eqn \ref{AUXMASTER1} there are two distinct sets of absorbing states; the first set contains only 
the fully extracted state $(N,0)$, the second set contains the states with only one hybridized base
$(i,N-i-1)$.  As we are seeking the probability of reaching the fully extracted state without having
any base rehybridize, we are only concerned with the $(N,0)$ absorbing state.  The probability
distribution for escape from the set of fully dehybridized states $(i,N-i)$ to the fully
extracted state $(N,0)$ is $S_{q}(t) = k_3^{(N-1)} W_{(N-1,1)}(t)$.  Then $Q_{i}(t,t^{*})$ is

\begin{equation}
\begin{array}{ccl}
  Q_{i}(t,t^{*}) & = & \int_0^{t^{*}-t} S_q(t)dt\\[13pt]
                 & = & k_3^{(N-1)} \int_0^{t^{*} - t} W_{(N-1,1)}(t;(i,N-i)) dt.\\[13pt]
\end{array}
\label{AUXMASTER2}
\end{equation}

\noindent Substituting Eqn \ref{AUXMASTER2}
into Eqn \ref{PATHPROB0}, and letting $t^{*} \rightarrow \infty$, we
have 

\begin{equation}
\begin{array}{ccl}
g_{i}(t) & = &  P_{i}(t) k_{1}^{(i)} k_3^{(N-1)} \int_0^{\infty} W_{(N-1,1)}(t;(i,N-i)) dt\\[13pt]
         &   &  + P_{i}(t) r_{1}^{(i)} k_3^{(N-1)} \int_0^{\infty} W_{(N-1,1)}(t;(i,N-i)) dt.\\[13pt]
\end{array}
\label{PATHPROB1}
\end{equation}

Note that in the limit of low target DNA concentrations, once the probe and target DNA have completely
dehybridized, rehybridization is unlikely during the time required to extract the probe DNA.
Thus, in the case of low target concentrations, the integral 
$k_3^{(N-1)} \int_0^{\infty} W_{(N-1,1)}(t;(i,N-i)) dt \approx 1$ and Eqn \ref{PATHPROB1} yields 
$g_{i}(t) = P_{(i,N-1-i)}(t)(r_{1}^{i} + k_{1}^{i})$.
With $g_{i}(t)$, we can compute the probability,
$\gamma_{i}$, that the last base to dehybridize 
is base $i$ from the expression 
$\gamma_{i} = \int_{0}^{\infty} g_{i}(t) dt$.  
The thermally averaged probability that
base $i$ is the last to dehybridize is computed from

\begin{equation}
\displaystyle \gamma_{i}^{\mathrm{(eq)}} = \frac{\sum_{n_1,n_2} \gamma_{i} e^{-\beta G(n_1,n_2) }}{\sum_{n_1,n_2} e^{-\beta G(n_1,n_2))}}.
\label{PATHPROB4}
\end{equation}

Referring to the inset in Fig \ref{BONDBREAK1}, we see that using the 
simple free energy model where each hybridized base lowers the free
energy of the system by $2kT$ produces a
$\gamma_i^{\mathrm{(eq)}}$ that is symmetric about the 
mid-point of the DNA lattice.  This is
exactly what is expected in the case where all
the attempt frequencies in the model
are equal and thermal 
denaturation dominates the dehybridization
process.  As the voltage increases, the 
behavior of $\gamma_i^{\mathrm{(eq)}}$ is also
predictable, with the bases at the $3'$ end of 
probe DNA becoming more likely to be the last to separate.
This is consistent with a voltage-driven 
denaturation mechanism where stripping of the
target DNA occurs.  Using the MFOLD free-energies for the 10c DNA 
segment, we see a similar effect where the
free energy surface dominates the last 
bond-probability at low voltages.  At higher
voltages the effects of the free energy surface 
are obscured and a stripping mechanism
dominates.  This is consistent with Fig \ref{LIFETIME1}, 
where we see that the mean first passage times are most strongly
influenced by sequence defects in the low voltage limit. 

This transition between a thermally dominated denaturation mechanism and a direct
stripping mechanism can explain the decreased influence of defects 
on the mean extraction time at high membrane voltages.
To demonstrate this, we consider the one-dimensional reduction of the
two-dimensional model of DNA dehybridization already presented.  Retaining
only the $5'$ hybridization transitions, the Master equation for the
one-dimensional model is

\begin{equation}
\begin{array}{llll}
\displaystyle \frac{dP(n_1,0)}{dt} & = & -(k_1(n_1) + r_1(n_1)(1-\delta_{n_1,0}))P(n_1,0) + &\\[13pt] 
                                   &   &  k_1(n_1 - 1)(1-\delta_{n_1,0})P(n_1 -1,0) +       &\\[13pt]
                                   &   &  r_1(n_1+1)(1-\delta_{n_1,N-1})P(n_1+1,0), &\\[13pt]
                                   &   &  \mathrm{   }0 \leq n_1 < N-1. &\\[13pt]
\end{array}
\label{MASTER1}
\end{equation}

The one-dimensional model of DNA denaturation described by Eqn \ref{MASTER1} is equivalent to
a one dimensional random walk with a reflecting boundary at $n_1 = 0$ and an absorbing boundary at
$n_1 = N$.  The first passage time for extraction into the transmembrane channel is then \cite{PURY}

\begin{equation}
\begin{array}{lll}
\displaystyle \tau(n_1)& = & \displaystyle \sum_{m=n_1}^{N-1} k_1(m)^{-1} + \\
& & \displaystyle \sum_{m=0}^{N-2} k_1(m)^{-1} \sum_{p=m+1}^{N-1} \prod_{j=m+1}^p \frac{r_1(j)}{k_1(j)} -\\
& & \displaystyle \sum_{m=0}^{n_1-2} k_1(m)^{-1} \sum_{p=m+1}^{n_1-1} \prod_{j=m+1}^p \frac{r_1(j)}{k_1(j)},\\
\label{FPRW1}
\end{array}
\end{equation}

\noindent and the thermally averaged escape time is

\begin{equation}
\displaystyle \tau_{\mathrm{eq}} = \frac{\sum_{n_1 = 0}^{N-1} \tau (n_1)e^{-\beta G(n_1,0)}}{\sum_{n_1 = 0}^{N} e^{-\beta G(n_1,0)}}.
\label{FPRW2}
\end{equation}

\noindent To make contact with the two dimensional model, we take the high voltage limit of Eqn \ref{FPRW1}.  Note that 
in the high voltage limit we expect the extraction times predicted by the one and two dimensional models to
agree to high accuracy.  Additionally, we focus on $\tau(0)$ as this is by far the dominant term in Eqn \ref{FPRW2}.
In the limit of high voltage $r_1(n_1)/k_1(n_1) = e^{\beta( (G(n_1 +1) - G(n_1)) - 2qV)} \rightarrow 0$ and

\begin{equation}
  \tau(0) = \sum_{m=0}^{N-1} \frac{1}{k_1(m)} = \sum_{m=0}^{N-1} \mu_1^{-1}e^{\beta(\Delta G(n_1) - qV)}.
\label{FPRW3}
\end{equation}

\noindent Here $\Delta G(n_1) = G(n_1 +1,0) - G(n_1,0)$.  Eqn \ref{FPRW3} shows that in the high
voltage limit, the mean extraction time is closely approximated by the 
time required to independently dehybridize $N$ DNA bases without significant rehybridization
occuring.  In this limit, a defect only influences the extraction
process once, when the defect base pair is first dehybridized.  Conversely, if we consider the
zero voltage limit, Eqn \ref{FPRW1} yields

\begin{equation}
\begin{array}{lll}
\displaystyle \tau(0) & = & \displaystyle \mu_1^{-1}e^{\beta(\Delta G(N-1) - qV)} +\\
& & \displaystyle + \sum_{m=0}^{N-2} \mu_1^{-1}e^{\beta(\Delta G(n_1) - qV)} \times \\
& & \left(1 + \sum_{p=m+1}^{N-1} \prod_{j=k+1}^{i} \left(\frac{\nu_1}{\mu_1}\right) e^{\beta(\Delta G(j))}\right),\\
\end{array}
\label{FPRW4}
\end{equation}

\noindent where the second term includes contributions from rehybridization events.
The effect of a defect is to reduce the ratio $r_1(j)/k_1(j) = (\nu_1/\mu_1) \e^{\beta(\Delta G(j))}$, and thus at low
voltages a defect accelerates the extraction process each time the defect base pair is denatured.  At low voltages
the base pair is typically denatured multiple times prior to probe exaction, and hence defects have a more pronounced
effect on the mean extraction time at low voltages than they do at higher voltages.  
We also note that the magnitude of Eqn \ref{FPRW4} depends strongly on the position
of the defect in the DNA sequence.  As was the case in the two-dimensional model, the mean extraction time
predicted by the one-dimensional model is a minimum when the defect is located in the middle of the target
strand.  As before, we ascribe this position dependence to a defect's ability to reduce 
the likelihood of rehybridization past the defect base pair, thus shortening the effective length of the
probe-target complex.

\section{CONCLUSION}

We have shown that the dehybridization process critical for the voltage-driven
transport of DNA through membrane channels can proceed through two distinct mechanisms.
In the low voltage regime, thermal dehybridization is the dominant mechanism and the details
of the hybridized DNA's free energy surface set the time scale for transport.  In the
moderate to high voltage regime, direct stripping of the complementary DNA off the probe
DNA is the dominant dehybridization mechanism.  Driven primarily by the applied voltage, this
dehybridization mechanism is comparatively insensitive to the shape of the hybridized 
DNA's free energy surface.  This behavior may have implications for the use of DNA sequencing
devices based on nanopores.  Specifically, an attempt to
increase the sequencing rate by increasing the voltage driving the double stranded DNA through the nanopore,
may obscure the distinction between different DNA sequences.  This may limit the
maximum rate at which sequencing can be performed.


\vspace{3mm}

GL and GNP acknowledge financial support from the Natural Science and Engineering
Research Council of Canada.  GL and TC acknowledge support from the National
Science Foundation through grant DMS-0206733, and the National Institutes of Health
through grant K25 AI058672.  The authors thank Andre Marziali, John Nakane, and
Matthew Wiggin for valuable comments on the manuscript.

\end{document}